\documentclass[12pt,preprint]{aastex}

\slugcomment{\mbox{\hspace{120pt}}Progress in Turbulence IX\\ 
\mbox{\hspace{120pt}}Springer Proceedings in Physics 267\\
\mbox{}
}

\shorttitle{Magnetoclinicity Instability}
\shortauthors{Yokoi \& Tobias}

\usepackage{graphicx}
\usepackage{dcolumn}
\usepackage{bm}

\usepackage{epstopdf}
\usepackage{oubraces}

\makeatletter
\input subeqn.sty
\input subeqna.sty
\input fancyhdr.sty
\makeatother

\begin{document}


\title{Magnetoclinicity Instability} 

\author{N. Yokoi\altaffilmark{1}}
\affil{Institute of Industrial Science, University of Tokyo,
    Tokyo 153-8505, Japan}
\email{nobyokoi@iis.u-tokyo.ac.jp}

\and

\author{Steven M. Tobias}
\affil{Department of Applied Mathematics, University of Leeds, Leeds LS2 9JT, UK}


\altaffiltext{1}{Visiting Researcher, Nordic Institute for Theoretical Physics (NORDITA).}


\begin{abstract}
In strongly compressible magnetohydrodynamic turbulence, obliqueness between the large-scale density gradient and magnetic field gives an electromotive force mediated by density variance (intensity of density fluctuation). This effect is named ``magnetoclinicity'', and is expected to play an important role in large-scale magnetic-field generation in astrophysical compressible turbulent flows. Analysis of large-scale instability due to the magnetoclinicity effect shows that the mean magnetic-field perturbation is destabilised at large scales in the vicinity of strong mean density gradient in the presence of density variance.
\end{abstract}





\section{Magnetoclinicity: Dynamo at strong compressibility\label{sec:1}}
With the aid of the two-scale direct-interaction approximation (TSDIA), a multiple-scale renormalised perturbation expansion theory for inhomogeneous turbulence \citep{yos1984,yok2020}, the turbulent electromotive force (EMF) is written as \citep{yok2018a,yok2018b}
\begin{equation}
	\langle {{\bf{u}}' \times {\bf{b}}'} \rangle
	= - (\beta + \zeta) \nabla \times {\bf{B}}
	+ \alpha {\bf{B}}
	- (\nabla \zeta) \times {\bf{B}}
	+ \gamma \nabla \times {\bf{U}}
	- \chi_{\rho} \nabla \overline{\rho} \times {\bf{B}}
	- \chi_Q \nabla Q \times {\bf{B}}
	- \chi_{D} \frac{D{\bf{U}}}{Dt} \times {\bf{B}},
	\label{eq:emf_tsdia}
\end{equation}
where ${\textbf{u}}'$ is the velocity fluctuation, ${\textbf{b}}'$ the magnetic fluctuation, ${\textbf{B}}$ the mean magnetic field, ${\textbf{U}}$ the mean velocity, $\overline{\rho}$ the mean density, $Q$ the mean internal energy, $D/Dt = \partial/\partial t + {\textbf{U} \cdot \nabla}$, and $\langle \cdots \rangle$ denotes ensemble averaging. Here, the transport coefficients $\eta_{\textrm{T}} (=\beta + \zeta)$, $\alpha$ and $\gamma$ represent the turbulent magnetic-diffusivity, residual-helicity and cross-helicity effects, respectively, which are present even in the incompressible case \citep{yok2013}. On the other hand, the transport coefficients $\chi_\rho$, $\chi_Q$, and $\chi_D$ have no counterparts in the incompressible case. They are related to the obliqueness of mean magnetic field to the gradients of density, internal energy, etc., and are called ``magnetoclinicity''. Note that in the TSDIA framework, they depend on the response functions and the compressible energy spectra with the multiplicational wavenumber factor $k^2$. This corresponds to the square of turbulent dilatation, $(\nabla \cdot {\textbf{u}}')^2$, and is directly connected to the magnitudes of density and internal-energy fluctuations. 

\begin{figure}[htpb]
\centering
\includegraphics[scale=1.0]{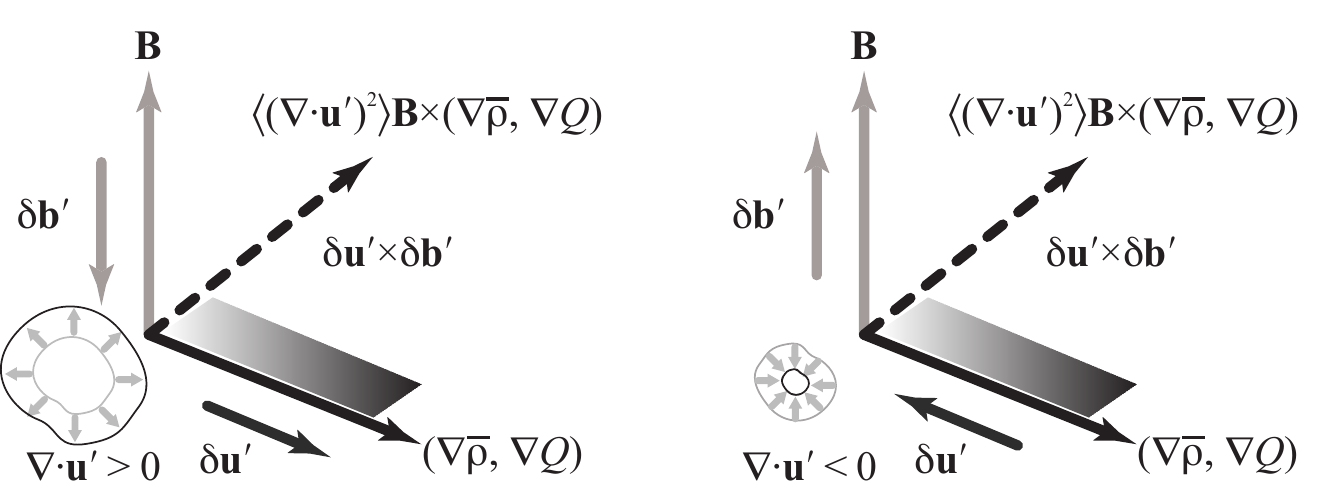}
\caption{Turbulent electromotive force due to the mis-alignment of the mean magnetic field ${\textbf{B}}$ from the gradient of mean density or internal energy, $\nabla \overline{\rho}$ or $\nabla Q$. Cases for the local expansion (positive dilatation) (left) and the local contraction (negative dilatation) (right).}
\label{fig:1}
\end{figure}

The physical origin of the magnetoclinicity effect can be obtained as follows. Through simplest linear relations, the density and internal-energy fluctuations can be expressed in terms of the turbulent dilatation as
\begin{equation}
	\rho' = - \tau_\rho \overline{\rho} 
  \nabla \cdot {\textbf{u}}',\;\;\;\;\;
	q' = -(\gamma_{\textrm{s}}-1) \tau_q Q \nabla \cdot {\textbf{u}}',
  \label{eq:rho_q_turb_dil}
\end{equation}
where $\gamma_{\textrm{s}}$ is the ratio of the specific heats at the constant pressure and volume, and $\tau_\rho$ and $\tau_q$ are the characteristic times for the density and internal-energy fluctuations, respectively. These relations naturally show that the density and internal-energy fluctuations are reduced or enhanced respectively with turbulent expansion ($\nabla \cdot {\textbf{u}}' >0$) or contraction ($\nabla \cdot {\textbf{u}}' <0$). From the equation of state, the fluctuation pressure is linearly related to the density and internal energy as $p' = \left( {\gamma_{\textrm{s}}-1} \right) 
  \left( {q' \overline{\rho} + \rho' Q} \right)$. Then the velocity fluctuation is related to the turbulent dilatation as
\begin{eqnarray}
	\frac{\partial{\textbf{u}}'}{\partial t}
	&=& \cdots - \frac{1}{\overline{\rho}} \nabla p' + \cdots
	\simeq \cdots - (\gamma_{\textrm{s}} -1) \frac{q'}{\overline{\rho}} 
		\nabla \overline{\rho}
	- (\gamma_{\textrm{s}} -1) \frac{\rho'}{\overline{\rho}} \nabla Q
	+ \cdots
	\nonumber\\
	&\simeq& \cdots + (\gamma_{\textrm{s}} -1)^2 
		\tau_q \frac{Q}{\overline{\rho}} 
		(\nabla \cdot {\textbf{u}}') \nabla \overline{\rho}
	+ (\gamma_{\textrm{s}} -1) 
		\tau_\rho (\nabla \cdot {\textbf{u}}') \nabla Q
	+ \cdots.
	\label{eq:fluct_u_eq}
\end{eqnarray}
Here, use has been made of (\ref{eq:rho_q_turb_dil}) on the final evaluation of (\ref{eq:fluct_u_eq}), which suggests that positive (negative) turbulent dilatation leads to velocity fluctuation parallel (anti-parallel) to the mean density gradient. On the other hand, from the induction equation of fluctuating magnetic field, we have
\begin{equation}
	\frac{\partial{\textbf{b}}'}{\partial t}
	= \cdots - (\nabla \cdot {\textbf{u}}') {\textbf{B}}
	+ \cdots.
	\label{eq:fluct_b_eq}
\end{equation}
This represents the effect of magnetoacoustic wave. Positive (negative) turbulent dilatation induces the magnetic fluctuation whose direction is opposite (parallel) to the mean magnetic field (Fig.~\ref{fig:1}).

Integrating (\ref{eq:fluct_u_eq}) and (\ref{eq:fluct_b_eq}) with respect to time, we get approximate expressions for ${\textbf{u}}'$ and ${\textbf{b}}'$. Then, the EMF due to turbulent dilatation, $\langle {{\textbf{u}}' \times {\textbf{b}}'} \rangle_{\textrm{TD}}$, is given as
\begin{eqnarray}
	\left\langle {
		{\textbf{u}}' \times {\textbf{b}}'
	} \right\rangle_{\textrm{TD}}
	\simeq &-& (\gamma_{\textrm{s}}-1)^2 
		\tau_u \tau_b \tau_q \left\langle {
		(\nabla \cdot {\textbf{u}}')^2
	} \right\rangle
	\frac{Q}{\overline{\rho}} 
	\nabla \overline{\rho} \times {\textbf{B}}
	\nonumber\\
	&-& (\gamma_{\textrm{s}}-1) 
	\tau_u \tau_b \tau_\rho \left\langle {
		(\nabla \cdot {\textbf{u}}')^2
	} \right\rangle
	\nabla Q \times {\textbf{B}},
	\label{eq:emf_td}
\end{eqnarray}
where $\tau_u$ and $\tau_b$ are the characteristic times of velocity and magnetic-field evolutions, respectively. Equation~(\ref{eq:emf_td}) infers that in the presence of the obliqueness between the mean magnetic field ${\textbf{B}}$ and the gradient of mean density, $\nabla \overline{\rho}$, and/or the gradient of mean internal energy, $\nabla Q$, the EMF is induced in the direction of ${\textbf{B}} \times \nabla \overline{\rho}$ and/or ${\textbf{B}} \times \nabla Q$, mediated by the turbulent dilatation. It is important to note that the direction of  $\langle {{\textbf{u}}' \times {\textbf{b}}'} \rangle_{\textrm{TD}}$ is always in the direction of ${\textbf{B}} \times \nabla \overline{\rho}$ and/or ${\textbf{B}} \times \nabla Q$, independent of the sign of turbulent dilatation (Fig.~\ref{fig:1}).

\section{Equilibrium state and disturbance \label{sec:2}}
In this work, we study a large-scale instability of compressible MHD turbulence: How do the mean or large-scale fields evolve under the influence of the turbulent transport represented by turbulent correlations such as the turbulent mass flux $\langle {\rho' {\textbf{u}}'} \rangle$, Reynolds stress $\langle {{\textbf{u}}'{\textbf{u}}'} \rangle$, turbulent Maxwell stress $\langle {{\textbf{b}}'{\textbf{b}}'} \rangle$, turbulent internal-energy flux $\langle {q' {\textbf{u}}'} \rangle$, EMF $\langle {{\textbf{u}}' \times {\textbf{b}}'} \rangle$, etc.\ appearing in the mean-field equations. For this purpose, a mean-field quantity $F$ is divided into the equilibrium unperturbed state $F_0$ and the deviation from it or disturbance, $\delta F$, as $F = F_0 + \delta F$ with the disturbance being much smaller than the equilibrium field: $|\delta F| \ll |F_0|$.

In this work, for the sake of simplicity, we assume simplified equilibrium mean fields for the velocity and magnetic field in the rectangular coordinate system $(x,y,z)$:
\begin{equation}
	{\textbf{U}}
	= \textbf{U}_0 + \delta {\textbf{U}}
	= \delta {\textbf{U}}
	= \left( {\delta U^x, \delta U^y, \delta U^z} \right),
	\label{eq:U_pert}
\end{equation}
\begin{equation}
	{\textbf{B}}
	= \textbf{B}_0 + \delta {\textbf{B}}
	= \left( {B_0,0,0} \right) 
	+ \left( {\delta B^x, \delta B^y, \delta B^z} \right).
	\label{eq:B_pert}
\end{equation}
The mean equilibrium velocity ${\textbf{U}}_0$ is assumed to be zero (${\textbf{U}}_0 = 0$), and the mean equilibrium magnetic field ${\textbf{B}}_0$ is put in the $x$ direction transverse to the mean equilibrium density gradient $\nabla \rho_0$ and uniform ($B_0 = {\textrm{const}}.$).

We decompose the mean-field equations into $F_0$ and $\delta F$ with (\ref{eq:U_pert}) and (\ref{eq:B_pert}), we have equations of disturbances:
\begin{equation}
	\frac{\partial \delta \rho}{\partial t}
	+ (\delta {\textbf{U}} \cdot \nabla) \rho_0
	+ \rho_0 \nabla \cdot \delta {\textbf{U}}
	= - \nabla \cdot \langle {\rho' {\textbf{u}}'} \rangle_1,
	\label{eq:delrho_eq}
\end{equation}
\begin{eqnarray}
	\lefteqn{
	\frac{\partial}{\partial t} \rho_0 \delta U^\alpha
	= - \frac{\partial \delta P}{\partial x^\alpha}
	+ \frac{\partial}{\partial x^a} 
    \mu \left( {
      \frac{\partial \delta U^\alpha}{\partial x^a}
      + \frac{\partial \delta U^a}{\partial x^\alpha}
	} \right)
	+ ({\textbf{J}}_0 \times \delta{\textbf{B}})^\alpha
	+ (\delta {\textbf{J}} \times {\textbf{B}}_0)^\alpha
	}\nonumber\\
	&&\hspace{-10pt}- \frac{\partial}{\partial x^a} \left[ {
		\delta\rho \left( {
			\langle {u'{}^a u'{}^\alpha} \rangle_0
			- \frac{1}{\mu_0 \rho_0} \langle {b'{}^a b'{}^\alpha} \rangle_0
		} \right)
		+ \delta U^a \langle {\rho' u'{}^\alpha} \rangle_0
		+ \delta U^\alpha \langle {\rho' u'{}^a} \rangle_0
	} \right],
	\label{eq:delU_eq}
\end{eqnarray}
\begin{eqnarray}
	\lefteqn{
	\frac{\partial}{\partial t} \left( {
		\rho_0 \delta Q + \delta\rho Q_0
	} \right)
	+ \nabla \cdot \left( {
		\rho_0 \delta{\textbf{U}} Q_0
	} \right)
	}\nonumber\\
	&&= \nabla \cdot \left( {
		\frac{\kappa}{C_v} \nabla \delta Q
	} \right)
	- \nabla \cdot \left( {
		\delta \overline{\rho} \langle {q' {\textbf{u}}'} \rangle_0
		+ \delta Q \langle {\rho' {\textbf{u}}'} \rangle_0
	} \right)
	\nonumber\\
	&&+ \delta{\textbf{U}} \langle {\rho' q'} \rangle_0
	- (\gamma_s -1) \left[ {
		\rho_0 Q_0 \nabla \cdot \delta{\textbf{U}}
		+ \delta\rho 
			\langle {q' \nabla \cdot {\textbf{u}}'} \rangle_0
		+ \delta Q \langle {\rho' \nabla \cdot {\textbf{u}}'} \rangle_0
	} \right],
	\label{eq:delQ_eq}
\end{eqnarray}
\begin{equation}
	\frac{\partial {\delta\textbf{B}}}{\partial t}
	= \nabla \times \left( {
		\delta{\textbf{U}} \times {\textbf{B}}_0
	} \right)
	+ \eta \nabla^2 \delta{\textbf{B}}
	+ \nabla \times \langle {{\textbf{u}}' \times {\textbf{b}}'} \rangle_1
	\label{eq:delB_eq}
\end{equation}
and the solenoidal condition of the magnetic field: $\nabla \cdot \delta{\textbf{B}} = 0$.

The pressure and internal-energy perturbations, $\delta P$ and $\delta Q$, can be expressed in terms of the density perturbation $\delta \rho$ with the speed of sound $c_{\textrm{s}}$ as
\begin{equation}
	\delta P
	=\left( {\gamma_{\textrm{s}} - 1} \right) \left( {
		\rho_0 \delta Q
		+ Q_0 \delta \rho
	} \right)
	= c_{\textrm{s}}^2 \delta\rho.
	\label{eq:pressure_int_en_cs_rel}
\end{equation}
Then, there is no need to solve the internal-energy equation.

The turbulent correlations in the mean-field perturbation equations are given as
\begin{equation}
	\langle {\rho' {\textbf{u}}'} \rangle_0
	= - \kappa_\rho \nabla \rho_0,\;\;\;\;\;
	\langle {\rho' {\textbf{u}}'} \rangle_1
	= - \kappa_\rho \nabla \delta \rho,
	\label{turb_mass_flux_pert}
\end{equation}
\begin{equation}
	\langle {u'{}^\alpha u'{}^\beta} \rangle_0
	- \frac{1}{\mu_0 \overline{\rho}} 
		\langle {b'{}^\alpha b'{}^\beta} \rangle_0
	= - \nu_{\textrm{K}} \left( {
		\frac{\partial U_0^\beta}{\partial x^\alpha}
		+ \frac{\partial U_0^\alpha}{\partial x^\beta}
	} \right)
	+ \nu_{\textrm{M}} \left( {
		\frac{\partial B_0^\beta}{\partial x^\alpha}
		+ \frac{\partial B_0^\alpha}{\partial x^\beta}
	} \right) \left/ {\mu_0 \overline{\rho}\rule{0.ex}{3.ex}} \right.
	= 0,
	\label{eq:rey_str_pert}
\end{equation}
\begin{equation}
	\langle {{\textbf{u}}' \times {\textbf{b}}'} \rangle_1
	= - \eta_{\textrm{T}} \delta {\textbf{J}}
	+ \alpha \delta{\textbf{B}}
	+ \gamma \delta \mbox{\boldmath$\Omega$}
	+ \chi_{\rho} {\textbf{B}}_0 \times {\nabla \delta\rho}
	+ \chi_{\rho} \delta {\textbf{B}} \times {\nabla \rho_0},
	\label{eq:emf_pert}
\end{equation}
where $\kappa_\rho$, $\nu_{\textrm{K}}$, and $\nu_{\textrm{M}}$ are the transport coefficients. Note that (\ref{eq:rey_str_pert}) gives no contribution because of the assumptions (\ref{eq:U_pert}) and (\ref{eq:B_pert}).

\section{Normal mode analysis of the mean-field equations \label{sec:3}}
We analyse an arbitrary disturbance into a complete set of normal modes, and examine the stability of each of these modes characterised by a wave number $k$. The disturbances are expressed in terms of two-dimensional periodic waves as
\begin{equation}
	\delta F 
	= \hat{f}(z) \exp[i(k^x x + k^y y) - i \omega_{\textbf{k}} t],
	\label{eq:normal_mode}
\end{equation}
where $\delta F = (\delta\rho, \delta{\textbf{U}}, \delta Q, \delta{\textbf{B}})$ and $\hat{f} = (\hat{\rho}, \hat{\textbf{u}}, \hat{q}, \hat{\textbf{b}})$. In general this formalism leads to a two-point boundary eigenvalue problem for the functions $\hat{f}(z)$. Here, as the simplest possible case, we assume that the amplitudes of disturbances, $\hat{f}$, do not depend on the vertical coordinate $z$ and constant, which will be relaxed in subsequent papers. Under this assumption, the equations of perturbations are
\begin{equation}
	\left( {
		- k^2 \kappa_\rho 
		+ i\omega_{\textbf{k}} 
	} \right) \hat{\rho}
	+ i k^x \rho_0 \hat{u}^x
	+ i k^y \rho_0 \hat{u}^y
	+ \frac{d\rho_0}{dz} \hat{u}^z
	= 0,
	\label{eq:hat_rho_eq}
\end{equation}
\begin{equation}
	- i k^x c_{\textrm{s}}^2 \hat{\rho}
	+ \left( {
		\kappa_\rho \frac{d^2 \rho_0}{dz^2}
		+ i\omega_{\textbf{k}} \rho_0
	} \right) \hat{u}^x
	= 0,
	\label{eq:hat_ux_eq}
\end{equation}
\begin{equation}
	- ik^y c_{\textrm{s}}^2 \hat{\rho}
	+ \left( {
		\kappa_\rho \frac{d^2 \rho_0}{dz^2}
		+ i \omega_{\textbf{k}} \rho_0
	} \right) \hat{u}^y
	- i k^y B_0 \hat{b}^x
	+ i k^x B_0 \hat{b}^y
	= 0,
	\label{eq:hat_uy_eq}
\end{equation}
\begin{equation}
	i k^x \kappa_\rho \hat{u}^x 
	+ i k^y \kappa_\rho \hat{u}^y
	+ \left( {
		\kappa_\rho \frac{d^2 \rho_0}{dz^2}
		+ i \omega_{\textbf{k}} \rho_0
	} \right) \hat{u}^z
	+ i k^x B_0 \frac{d\rho_0}{dz} \hat{b}^z
	= 0,
	\label{eq:hat_uz_eq}
\end{equation}
\begin{equation}
	k^2 \gamma \hat{u}^x 
	- i k^y B_0 \hat{u}^y
	+ \left( {
		- k^2 \eta_{\textrm{T}}
		+ \chi_\rho \frac{d^2\rho_0}{dz^2}
		+ i\omega_{\textbf{k}}
	} \right) \hat{b}^x
	+ i k^y \alpha \hat{b}^z
	= 0,
	\label{eq:hat_bx_eq}
\end{equation}
\begin{equation}
	\left( {
		k^2 \gamma + i k^x B_0
	} \right) \hat{u}^y
	+ \left( {
		- k^2 \eta_{\textrm{T}}
		+ \chi_\rho \frac{d^2\rho_0}{dz^2}
		+ i\omega_{\textbf{k}}
	} \right) \hat{b}^y
	- i k^x \alpha \hat{b}^z
	= 0,
	\label{eq:hat_by_eq}
\end{equation}
\begin{equation}
	i k^x B_0 \hat{u}^x
	+ k^2 \gamma \hat{u}^z
	- i k^y \alpha \hat{b}^x
	+ i k^x \alpha \hat{b}^y 
	+ \left( {
		- k^2 \eta_{\textrm{T}}
		+ i\omega_{\textbf{k}}
	} \right) \hat{b}^z
	= 0.
	\label{eq:hat_bz_eq}
\end{equation}
This system of equations (\ref{eq:hat_rho_eq})-(\ref{eq:hat_bz_eq}) with the solenoidal conditions for the magnetic field is analysed. One of the dispersion relations is given by
\begin{equation}
	\chi_\rho \frac{d^2 \rho_0}{dz^2} 
	- \eta_{\textrm{T}} k^2 
	+ i \omega_{\textbf{k}}
	=0.
	\label{eq:dispersion_rel}
\end{equation}
From this, the $\alpha$ component of large-scale magnetic-field disturbance is written as
\begin{equation}
	\delta B^\alpha 
	=  \hat{b}^\alpha \exp \left[ {
		\left( {- \eta_{\textrm{T}} k^2
		+ \chi_\rho \frac{d^2 \rho_0}{dz^2}} \right) t
	} \right]
	\exp[i(k^x x + k^y y)].
	\label{eq:mag_pert_evol}
\end{equation}
The first term in the temporal evolution part arises from the turbulent magnetic diffusivity $\eta_{\textrm{T}}$. The growth of the mean-field perturbations are suppressed by $\eta_{\textrm{T}}$. This effect is strongest at small scales where the wave number $k$ is large. On the other hand, in the presence of a strong mean density inhomogeneity such that
\begin{equation} 
	\chi_\rho \frac{d^2 \rho_0}{dz^2} > \eta_{\textrm{T}} k^2,
	\label{eq:destab_cond}
\end{equation}
the second or $\chi_\rho$-related term in the temporal evolution part contributes to the growth of mean-field perturbations. This large-scale instability, the magnetoclinicity instability, is important only in the region where the density variance is strong enough since it also depends on $\chi_\rho (\propto \langle {\rho'{}^2} \rangle)$.

\section{Instability across the strong density variation \label{sec:4}}
\begin{figure}[htb]
\centering
\includegraphics[scale=1.25]{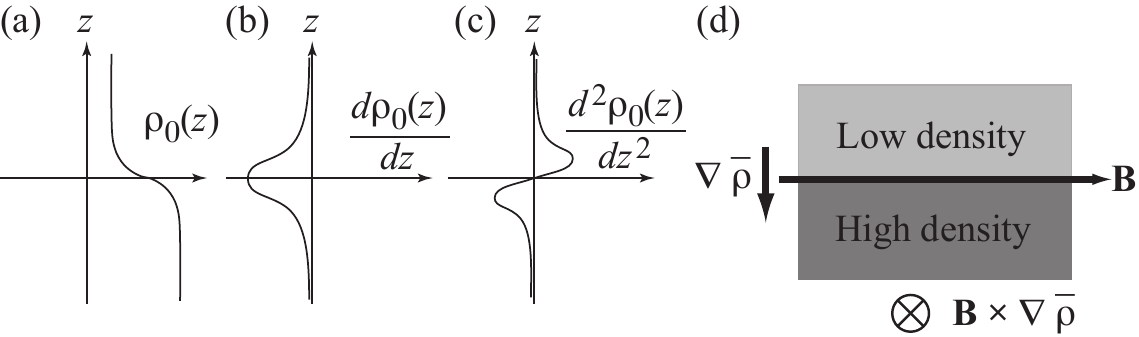}
\caption{Schematic spatial distributions of (a) the unperturbed density $\rho_0(z)$, (b) its first derivative with respect to $z$, $d^2 \rho_0/dz$, (c) the second derivative $d^2 \rho_0/dz^2$, and (d) the setup with transverse ${\textbf{B}}$.}
\label{fig:2}
\end{figure}

In order to quantitatively evaluate the magnetoclinicity effect, we consider a simplest possible spatial profile of the unperturbed density $\rho_0(z)$ as
\begin{equation}
	\rho_0(z)
	= \rho_{\textrm{m}} - \rho_{\textrm{d}} \tanh 
		\left( {{z}/{z_{\textrm{d}}}} \right),
	\label{eq:unpert_den_profile}
\end{equation}
where $\rho_{\textrm{m}} [= (\rho_> + \rho_<)/2]$ is the reference (average) density, $\rho_{\textrm{d}} [= (\rho_> - \rho_<)/2]$ the density difference, and $z_{\textrm{d}}$ the depth of mean density variation. For the spatial distribution of unperturbed density (\ref{eq:unpert_den_profile}), the first and second derivatives are given as
\begin{equation}
	\frac{d\rho_0(z)}{dz}
	= - \frac{\rho_{\textrm{d}}}{z_{\textrm{d}}} 
		\frac{1}{\cosh^2 \left( {{z}/{z_{\textrm{d}}}} \right)},\;\;\;\;\;
	\frac{d^2 \rho_0(z)}{dz^2}
	= + \frac{2\rho_{\textrm{d}}}{z_{\textrm{d}}^2} 
	\frac{\tanh(z/z_{\textrm{d}})}{\cosh^2 
		\left( {{z}/{z_{\textrm{d}}}} \right)}.
	\label{eq:1st_2nd_der_rho}
\end{equation}
The schematic spatial distribution of the unperturbed density, its first and second derivatives, as well as the setup considered, are depicted in Fig.~\ref{fig:2}.

With this density configuration, the second derivative is positive in the upper layer (low density region) and negative in the lower layer (high density region) as
\begin{equation}
	\frac{d^2 \rho_0}{dz^2}
	\left\{ {
		\begin{array}{ll}
		\rule{0.ex}{2.ex}
		>0 & (z>0,\; \rho_<:\; \mbox{low density}),\\
		\rule{0.ex}{2.ex}
		<0 & (z<0,\; \rho_>:\; \mbox{high density}).
		\end{array}
	} \right.
	\label{eq:2nd_der_rho_signs}
\end{equation}
It follows from (\ref{eq:mag_pert_evol}) that the mean magnetic-field disturbance can increase in the low density (positive $z$) side, and decays in the high-density (negative $z$) side.  The lower the wave number $k$ is, the larger the growth rate of the perturbed magnetic field is. In this sense, this magnetoclinicity effect is more suitable for producing large-scale magnetic-field structures than small-scale ones. The growth rate also depends on how much large transport coefficient $\chi_\rho$ is. The magnitude of $\chi_\rho$ reflects the magnitude of density variance $\langle {\rho'{}^2} \rangle$. If the high $\chi_\rho$ region is spatially localised, the instability region of the large magnetic field is also spatially localised. A region with a strong mean density gradient $\nabla \overline{\rho}$ is favourable for high density variance $\langle {\rho'{}^2} \rangle$, since $\langle {\rho'{}^2} \rangle$ is generated by strong $\nabla \overline{\rho}$ coupled with $- \langle {\rho' {\textbf{u}}'} \rangle$. We stress again here that although the arguments here make physical sense, a global analysis involving a two-point boundary value problem is necessary to elucidate the mechanisms.

\acknowledgements
{This work was supported by Japan Society for the Promotion of Science (JSPS) Grants-in-Aid for Scientific Research: JP18H01212.}
%



\begin{thebibliography}{99}%


\bibitem[Yoshizawa(1984)]{yos1984} 
A Yoshizawa (1984) Statistical analysis of the deviation of the Reynolds stress from its eddy-viscosity representation, Phys Fluids 27:1377--1387

\bibitem[Yokoi(2020)]{yok2020} 
N Yokoi (2020) Turbulence, Transport and Reconnection, in D. MacTaggart and A. Hillier (eds.), {\it Topics in Magnetohydrodynamic Topology, Reconnection and Stability Theory}, CISM International Centre for Mechanical Sciences 591, Springer: 177--265

\bibitem[Yokoi(2018a)]{yok2018a} 
N Yokoi (2018a) Electromotive force in strongly compressible magnetohydrodynamic turbulence, J Plasma Phys 84:735840501-1--26 

\bibitem[Yokoi(2018b)]{yok2018b} 
N Yokoi (2018b) Mass and internal-energy transports in strongly compressible magnetohydrodynamic turbulence, J Plasma Phys 84:7758140603-1--30

\bibitem[Yokoi(2013)]{yok2013} 
N Yokoi (2013) Cross helicity and related dynamo, Geophys Astrophys Fluid Dyn 107:114-184

\end{thebibliography}
\end{document}